# Bitcoin price and its marginal cost of production: support for a fundamental value

**Abstract** This study back-tests a marginal cost of production model proposed to value the digital currency bitcoin. Results from both conventional regression and vector autoregression (VAR) models show that the marginal cost of production plays an important role in explaining bitcoin prices, challenging recent allegations that bitcoins are essentially worthless. Even with markets pricing bitcoin in the thousands of dollars each, the valuation model seems robust. The data show that a price bubble that began in the Fall of 2017 resolved itself in early 2018, converging with the marginal cost model. This suggests that while bubbles may appear in the bitcoin market, prices will tend to this bound and not collapse to zero.

Adam S. Hayes, CFA [1] [*]
1 Department of Sociology, University of Wisconsin-Madison, Madison, WI, USA; and
  MSc Program in Digital Currencies, University of Nicosia, Cyprus

* Corresponding author
E-mail: ahayes8@wisc.edu
Ph: +1 917-494-5475

# Bitcoin price and its marginal cost of production: support for a fundamental value

**Abstract** This study back-tests a marginal cost of production model proposed to value the digital currency bitcoin. Results from both conventional regression and vector autoregression (VAR) models show that the marginal cost of production plays an important role in explaining bitcoin prices, challenging recent allegations that bitcoins are essentially worthless. Even with markets pricing bitcoin in the thousands of dollars each, the valuation model seems robust. The data show that a price bubble that began in the Fall of 2017 resolved itself in early 2018, converging with the marginal cost model. This suggests that while bubbles may appear in the bitcoin market, prices will tend to this bound and not collapse to zero.

**1. Introduction**

The price of bitcoin increased from $250 in June of 2015 to more than $19,000 by December of 2017, where it commanded a market capitalization in excess of $325 billion before stabilizing around the $10,000 level. [1] Presently, Bitcoin sees many hundreds of millions of dollars' worth of transactions cross its system on a daily basis; and yet, the cryptocurrency functions entirely as its own decentralized computer network, without any central bank, government, or regulatory body to back it. This has led some to conclude that the price of the cryptocurrency is a massive speculative bubble, with some researchers claiming that there is *no* fundamental underpinning to its value (e.g. Hanley 2013; Yermack 2013). Cheah and Fry (2015), among others, echo recent comments made by Jamie Dimon, the CEO of investment bank JPMorgan Chase, in asserting that the fundamental value of bitcoin is indeed zero, and that the entire pursuit is a fool's errand, or worse a fraud.[2] Even the Wall Street Journal has opined that Bitcoin is "probably worth zero."[3] Nonetheless, the popularity of the cryptocurrency continues to rise, and its price remains far from nil. Against this backdrop, it is of growing concern to evaluate the basis for value of bitcoin.

---

[1] For consistency, Bitcoin with a capital 'B' refers to the general system, network and protocol, while bitcoin with a small 'b' refers to the digital currency itself or units thereof
[2] https://www.bloomberg.com/news/articles/2017-09-12/jpmorgan-s-ceo-says-he-d-fire-traders-who-bet-on-fraud-bitcoin
[3] https://www.wsj.com/articles/bitcoins-wild-ride-shows-the-truth-it-is-probably-worth-zero-1505760623

Challenging the views of Mr. Dimon and the grim hypotheses of some skeptical researchers, Hayes (2016) suggests that bitcoin does indeed have a quantifiable intrinsic value and formalizes a pricing model based on its marginal cost of production:[4] "mining," or the process of creating new bitcoins through concerted computational effort requires the consumption of electric power, which incurs a real monetary cost for mining participants, and thus the value of bitcoin is the embodied costs of production (on the margin).

This study seeks to test the validity of this cost of production theory of value by back-testing the pricing model against the observed market price, going back nearly five years. A simple OLS regression indicates that the model price explained approximately 81% of the observed market price and a striking 97% of the observed changes in market prices over that period. Following this up, a Granger test on the postestimation results of a subsequent vector autoregression (VAR) model is carried out, which strongly rejects the null hypothesis that the pricing model does not "cause" the market price. The Granger test is used here not assert causality, but to support the notion that the modeled price and the observed price match up to a statistically significant degree over time.

**2. The Cost of Production Model**

The process and technical elaboration of bitcoin production ("mining") is described at length elsewhere (e.g. Kroll et al. 2013; Sapirshtein et al. 2016; Nakamoto 2008). Suffice it to say that mining involves a competition among producers; with a novel feature that the rate of new unit formation is *fixed* so that increased demand cannot induce a greater supply, and so this elasticity is manifest instead through increased difficulty in the production process itself, increasing the system-wide marginal cost of production.

---

[4] Or at least, an expected lower bound to its market price

The primary ongoing cost for bitcoin production is that of electricity, measured in dollars per kilowatt-hour (kWh). Of course, different regions of the world will consume electricity at their local rates (which may vary by customer type, power generation source, and time of day) and in their local currencies, but for the sake of convenience it is a good working assumption that the average rate of electricity worldwide accounting for both residential and commercial rates is approximately USD $0.135 per kWh.[5]

Following Hayes (2016: ), The (marginal) cost of production per day, $E_{day}$ per unit of mining power can be expressed as:

$$E_{day} = (\rho/1000)(\$/kWh \cdot WperGH/s \cdot hr_{day}) \quad (1)$$

where: $E_{day}$ is the dollar cost per day for a producer, $\rho$ is the hashpower (computational power) employed by a miner, $/ kWh is the dollar price per kilowatt-hour, and W per GH/s is the energy efficiency of the hardware, and $hrs_{day}$ is the number of hours in a day.

In order to calculate the expected number of bitcoins the same miner can produce daily, the following equation is used to calculate the daily (marginal) product:

$$BTC/day^* = \left(\frac{\beta\rho \cdot \sec_{hr}}{\delta \cdot 2^{32}}\right) hr_{day} \quad (2)$$

where: BTC/day* is the expected level of daily bitcoin production when mining bitcoin, β is the block reward (expressed in units of BTC/block), ρ is the hashing power employed by a miner, and δ is the difficulty (expressed in units of GH/block) The constant $sec_{hr}$ is the number of seconds in an hour, $hr_{day}$ the number of hours in a day. Presently, the block reward is 12.5 BTC per block.

---

[5] https://en.wikipedia.org/wiki/Electricity_pricing

According to microeconomic theory, under conditions of competition, the marginal product should equate with its marginal cost, which should also equal its selling price. Because of this theoretical equivalence, and since cost per day is expressed in terms of $/day and production in BTC/day, the $/BTC price level is revealed as the ratio of (cost/day) divided by (BTC/day). This objective price of production, $P^*$, serves as a logical lower bound for the market price, below which a producer would operate at a marginal loss and presumably remove themselves from the network. $P^*$ is expressed in dollars per bitcoin, given the difficulty and cost of production:

$$P^* = \frac{E_{day}}{BTC/day^*} \qquad (3)$$

**3. Testing the Model Empirically**

I back-test the above model using historical observed price data and compare that to what the model would have predicted. Observed market price and difficulty data were collected using the website blockchain.info, a reliable and transparent source of Bitcoin market and protocol data, at the dates of difficulty changes in the network (approximately once every two weeks) to consistently measure market price given a particular value of mining difficulty, from June 29, 2013 through April 27, 2018.

The model requires as an input the average energy efficiency of the mining network. This information was extracted from Bitcoin mining hardware manufacturer websites and checked against a dedicated wiki page that catalogues the efficiency of current mining hardware (https://en.bitcoin.it/wiki/Mining_hardware_comparison). This data was then collected for each date of difficulty change, scraped from the web using the internet archive's *wayback machine* (https://web.archive.org/web/20170215000000*/https://en.bitcoin.it/wiki/Mining_hardware_comparison). As stated above, for simplicity, I hold electricity costs constant at 13.5 cents per kilowatt-hour. Table A1, which appears in the Appendix, describes these data points along with estimated model price for each difficulty change date.

### 3.1 Conventional Regression Analysis

As a first pass, I compared the ratio of observed price to modeled price over time, from June 2013 through April 2018. As Figure 1 shows, since June 2013, the market price has tended to fluctuate about the price estimated by the model. In the chart, a y-axis value of 1.00 indicates that the market price and model price are identical. Values above 1.00 indicate a premium in the market relative to the model and below 1.00 a relative discount. Over the long-run (~5 years), the average ratio is 1.05, σ = 0.33, which is striking in its accuracy. This suggests that the market for bitcoin has been quite efficient from a production standpoint, if not volatile, contradicting assertions that this market is consistently inefficient (e.g. Urquhart 2016). There is evidence of increased volatility from approximately September 2017 through January 2018, indicating that the market had deviated substantially from the model, but did eventually converge once again. This spike indicates the emergence and reconciliation of a price bubble; however, the presence of such bubbles does not indicate a zero value, only that prolonged departures from the modeled price can exist, but which ultimately resolve to the marginal cost of production.

*Figure 1*

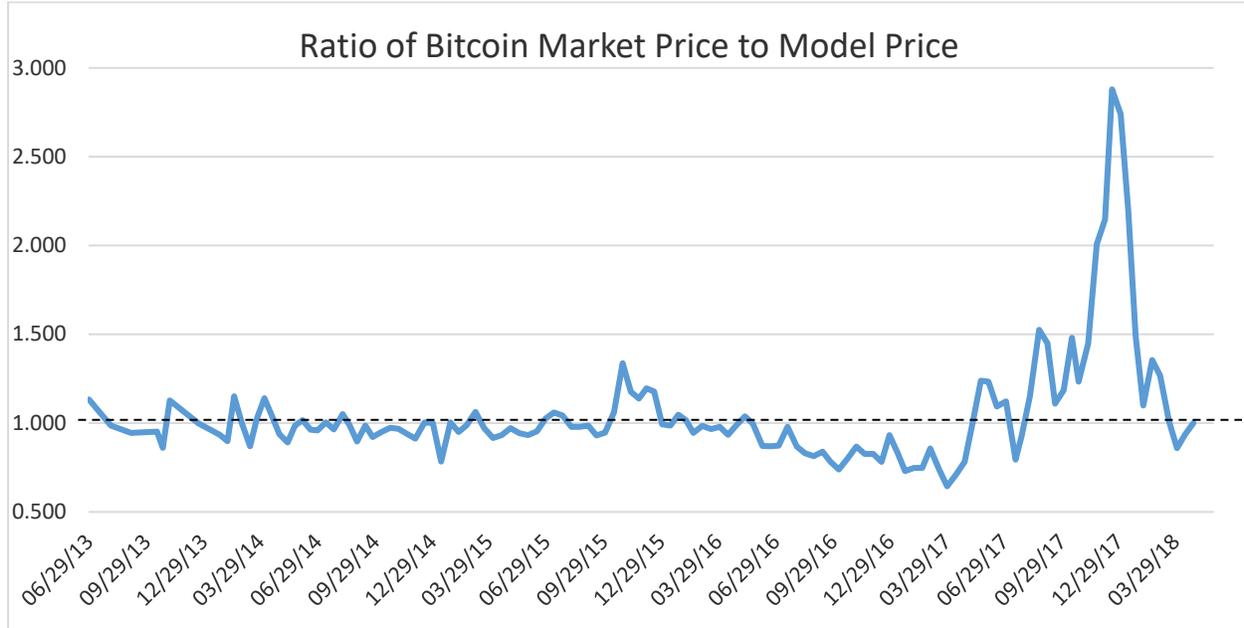

*Figure 1:* Ratio of bitcoin price observed in the market to the expected price produced by the model using historical data (source: www.blockchain.info). 1.00 would indicate that the two prices are identical, anything over 1.00 indicates a premium in the market and below a discount. The average for the study period is 1.05, σ = 0.33, indicating that over the long-term, the market price seems to fluctuate around the modeled price with striking consistency.

This initial result is suggestive, and so a more rigorous analysis was undertaken to test the "fit" of the pricing model against observed historical data. The valuation model output and observed prices appear in Figure 2, with what amounts to two time series for comparison. A conventional OLS regression was first carried out to obtain a proxy for model fit and to judge how much of the market price is described by the model; which produces $R^2 = 0.813$, telling us that 81% of the observed market price can be explained by the marginal cost of production model over the sample period. Next, I conduct a second OLS regression on the log transformations of each time series, yielding an $R^2 = 0.969$, suggesting that nearly all of the marginal change in market price can be explained by the change in marginal cost.

*Figure 1*: Historical bitcoin market price vs. ex-post implied model price, June 2013-April 2018

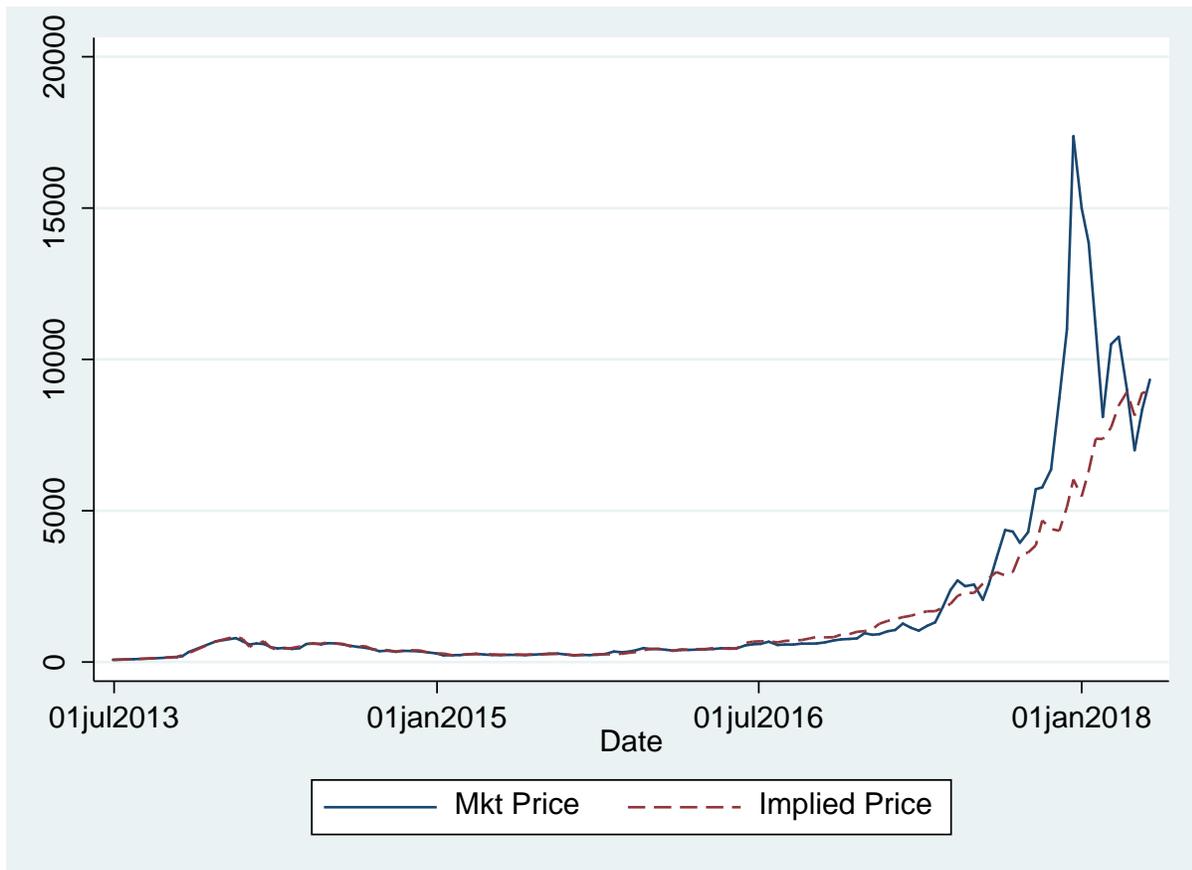

### 3.2 VAR Granger Analysis

Next, in order to compare these two time series directly to each other in a methodologically rigorous way, I estimate a multivariate vector autoregression (VAR) with two lags each on the log transformation of market price and implied model price.[6] The purpose of the VAR is primarily to test the postestimation results using a Granger test (Geweke 1982). Typically used to suggest temporal causality, I instead use this test here to evaluate the post-hoc predictive power of the cost of production pricing model. The test considers two null hypotheses:

---

[6] Testing for autocorrelation suggests that two is the appropriate number of lags.

$H_01$: The market price does *not* "cause" the model price; and

$H_02$: The model price does *not* "cause" the market price.

As Table 1 shows, $H_01$ cannot be rejected, which is to be expected: the model is supposed to describe the market and not the other way around. $H_02$ however, is strongly rejected, and the alternative hypothesis that the model price implies the market price is given a large degree of support ($p < 0.001$). This key finding lends credibility that the marginal cost of production of bitcoin describes its price and disputes those who claim that bitcoin is worthless.

**Table 1: Granger Test on VAR Postestimation Equations**

Granger Causality Wald Tests

| H0 | chi2 | df | Prob >chi2 |
|---|---|---|---|
| 1: Market price does not Granger cause the model price | 4.579 | 2 | 0.101 |
| 2: Model price does not Granger cause the market price | 13.301 | 2 | 0.001 |

*Note*: H01 cannot be rejected, which is to be expected: the model is supposed to describe the market and not the other way around. H02 however, is strongly rejected, and the alternative hypothesis that the model price implies the market price is given a large degree of support ($p < 0.001$). This key finding lends credibility that the marginal cost of production of bitcoin describes its price and disputes those who claim that bitcoin is worthless.

**4. Discussion & Conclusion**

The marginal cost of production has been proposed as a model to value bitcoin (Hayes 2016). In this paper, the cost of production model was back-tested using historical data showing that the market price of bitcoin tends to fluctuate around the model price, and with the model price explaining the market price in a statistically significant manner.

This finding is striking given the volume of recent media accounts and research projects that have supposed no fundamental value at all for bitcoin (e.g. Cheah and Fry 2015). Moreover, it suggests that attempts to find a causal link between the long-run price of bitcoin and various exogenous factors may be misguided (e.g. Ciaian et al. 2016; Kristoufek 2015; Polasik et al. 2015), as well as attempts to value bitcoin as if it were a traditional financial asset (e.g. Cretarola et al. 2017).

These findings are also indicative that the bitcoin market is susceptible to price bubbles, as has been suspected. However, despite a significant deviation in price to the upside from the Fall of 2017 through early 2018, the cost or production model has remained resilient as the market price did ultimately converge with the model. This novel pricing method leads us to expect that during periods of excess demand (e.g. a price bubble), either the market price will fall and/or the mining difficulty will increase to resolve the discrepancy. In the case of the late 2017 bubble just described, it does appear that both mechanisms were at play: the price fell and the mining difficulty rose simultaneously. Bubbles in the bitcoin market have been explored in-depth elsewhere (e.g. Garcia et al. 2014; Cheah and Fry 2015; Li et al. 2018; Hafner 2018). Cheung et al. (2015) as well as Su et al. (2018) use the Phillips-Shi-Yu (2013, 2015) method of bubble detection, confirming that multiple short-lived bubbles have characterized bitcoin prices, with four "explosive bubbles" since 2011 inclusive of the late-2017 period already described (see also: Corbet et al. 2017). The current study adds to this literature suggesting that while bubbles can – and indeed have – existed in the bitcoin market, the resolution of such a bubble will not be a collapse toward zero, but rather toward its marginal cost of production. Future analysis of bitcoin bubbles given this prediction can be explored further following the method elaborated by Pavlidis et al. (2017) since the recent introduction of bitcoin futures markets in December 2017. This type of analysis, however, will require a lengthier time series than what is presently available.

It is important to note that the above analyses apply primarily to bitcoin and does not necessarily extend to other cryptocurrencies such as Ethereum or Litecoin; although a similar study may indeed support the cost of production thesis there as well. Still, with Bitcoin dominating the digital currency market, both in scale and scope, it is a worthwhile pursuit to understand why this unique asset has value.

**References**

1. Cheah ET, Fry J. Speculative bubbles in Bitcoin markets? An empirical investigation into the fundamental value of Bitcoin. Economics Letters. 2015 May 31;130:32-6.

2. Cheung A, Roca E, Su JJ. Crypto-currency bubbles: an application of the Phillips–Shi–Yu (2013) methodology on Mt. Gox bitcoin prices. Applied Economics. 2015 May 15;47(23):2348-58.

3. Ciaian, P., Rajcaniova, M., & Kancs, D. A. The economics of Bitcoin price formation. Applied Economics. 2016 48(19):1799-1815.

4. Corbet S, Lucey B, Yarovya L. Datestamping the Bitcoin and Ethereum bubbles. Finance Research Letters. 2017 Dec 19.

5. Cretarola A, Figà-Talamanca G, Patacca M. A sentiment-based model for the BitCoin: theory, estimation and option pricing. arXiv preprint arXiv:1709.08621. 2017 Sep 23.

6. Garcia D, Tessone CJ, Mavrodiev P, Perony N. The digital traces of bubbles: feedback cycles between socio-economic signals in the Bitcoin economy. Journal of the Royal Society Interface. 2014 Oct 6;11(99):20140623.

7. Geweke, J. Measurement of linear dependence and feedback between multiple time series. Journal of the American statistical association, 1982. 77(378), 304-313.

8. Hafner, C.,Testing for bubbles in cryptocurrencies with time-varying volatility. 2018

9. Hanley BP. The false premises and promises of Bitcoin. arXiv preprint arXiv:1312.2048. 2013.

10. Hayes AS. Cryptocurrency Value Formation: An empirical study leading to a cost of production model for valuing Bitcoin. Telematics and Informatics. 2016.

11. Kristoufek, L. What are the main drivers of the Bitcoin price? Evidence from wavelet coherence analysis. PloS one. 2015 10(4):e0123923.

12. Kroll, J. A., Davey, I. C., & Felten, E. W. (2013, June). The economics of Bitcoin mining, or Bitcoin in the presence of adversaries. In Proceedings of WEIS (Vol. 2013)

13. Li ZZ, Tao R, Su CW, Lobonţ OR. Does Bitcoin bubble burst?. Quality & Quantity. 2018:1-5.

14. Nakamoto, S.. Bitcoin: A peer-to-peer electronic cash system. 2008

15. Pavlidis EG, Paya I, Peel DA. Testing for speculative bubbles using spot and forward prices. International Economic Review. 2017 Nov 1;58(4):1191-226.

16. Phillips RC, Gorse D. Cryptocurrency price drivers: Wavelet coherence analysis revisited. PloS one. 2018 Apr 18;13(4):e0195200.

17. Phillips PC, Shi S, Yu J. Technical supplement to the paper: Testing for multiple bubbles: Limit theory of real time detectors. Manuscript, available from https://sites. google. com/site/shupingshi/TN_GSADF. pdf. 2013.

18. Phillips PC, Shi S, Yu J. Testing for multiple bubbles: Historical episodes of exuberance and collapse in the S&P 500. International Economic Review. 2015 Nov 1;56(4):1043-78.

19. Polasik, M., Piotrowska, A. I., Wisniewski, T. P., Kotkowski, R., & Lightfoot, G. Price fluctuations and the use of Bitcoin: An empirical inquiry. International Journal of Electronic Commerce, 2015 20(1):9-49.

20. Sapirshtein, A., Sompolinsky, Y., & Zohar, A. Optimal selfish mining strategies in bitcoin. In International Conference on Financial Cryptography and Data Security. 2016. (pp. 515-532). Springer, Berlin, Heidelberg.

21. Su CW, Li ZZ, Tao R, Si DK. Testing for multiple bubbles in bitcoin markets: A generalized sup ADF test. Japan and the World Economy. 2018 Mar 23.

22. Urquhart A. The inefficiency of Bitcoin. Economics Letters. 2016 Nov 30;148:80-2.

23. Yermack D. Is Bitcoin a real currency? An economic appraisal. National Bureau of Economic Research; 2013 Dec 19.